\documentclass[preprint,showpacs,floatfix]{revtex4-1}
\usepackage{amssymb}
\usepackage{amsmath}
\usepackage[dvips]{graphicx}

\def\be{\begin{eqnarray} &&}

\def\ee{\end{eqnarray}}

\def\bew{\begin{widetext}}
\def\ew{\end{widetext}}

\def\Dslash{\raise.15ex\hbox{/}\kern-.7em D}
\def\Pslash{\raise.15ex\hbox{/}\kern-.7em P}

\newcommand{\bge}{\begin{equation}}
\newcommand{\ene}{\end{equation}}
\newcommand{\bea}{\begin{eqnarray}}
\newcommand{\eea}{\end{eqnarray}}
\newcommand{\bg}{\begin{eqnarray}}
\newcommand{\en}{\end{eqnarray}}

\def\qbar{\bar{q}}
\def\Qbar{\overline{Q}}

\usepackage{color}

\renewcommand{\bar}[1]{\overline{#1}}

\begin{document}
\title{ Pion structure in the nuclear medium}   
\author{
J.~P.~B.~C.~de Melo$^a$\footnote{joao.mello@cruzeirodosul.edu.br},
K.~Tsushima$^b$\footnote{kazuo.tsushima@gmail.com},
Bruno~El-Bennich$^a$\footnote{{bruno.elbennich@cruzeirodosul.edu.br}}~,
E.~Rojas$^a$\footnote{eduardo.rojas@cruzeirodosul.edu.br}
and
T.~Frederico$^c$\footnote{tobias@ita.br}
}
\affiliation{\
$^a$Laborat\'orio de F\'\i sica Te\'orica e Computacional\\
Universidade Cruzeiro do Sul\\
01506-000, S\~ao Paulo, SP, Brazil\\
$^b$International Institute of Physics\\
Federal University of Rio Grande do Norte\\
59078-400, Natal, RN, Brazil\\
$^c$Instituto Tecnol\'ogico da Aeron\'autica/DCTA\\
12228-900, S\~ao Jos\'e dos Campos, SP, Brazil
}
\date{\today}

\begin{abstract}

Using the light-front pion wave function based on a Bethe-Salpeter amplitude model,
we study the properties of the pion in symmetric nuclear matter. The pion model we adopt
 is well constrained by previous studies to explain the pion properties in vacuum.
In order to consistently incorporate the constituent up and down quarks of the pion
immersed in symmetric nuclear matter, we use the quark-meson coupling model, which
has been widely applied to various hadronic and nuclear phenomena
in a nuclear medium with success. We predict the in-medium modifications of
the pion electromagnetic form factor,
charge radius and weak decay constant in symmetric nuclear matter.

\end{abstract}


\maketitle

\section{Introduction}
\label{intro}

One of the most exciting challenges in hadronic physics
is to investigate the changes in hadron properties in
a nuclear medium as well as in a nucleus~\cite{Hayano,Medium1}.
In particular, partial restoration of chiral symmetry in a dense
nuclear medium has not yet been confirmed by experiment, although
it is generally accepted to occur based on quantum chromodynamics (QCD).
In spite of the notorious complex number or sign problem of the fermion
determinant, one hopes that lattice QCD simulation will eventually be able
to study hadron properties in a nuclear medium with relatively high baryon
densities~\cite{Lattice1,Lattice2,Lattice3,Lattice4}.

Chiral symmetry is one of the most important symmetries in QCD. Therefore, it is very interesting to study the pion properties,
since it is the  Nambu-Goldstone boson of the theory which is realized in nature due to the spontaneous breaking of chiral symmetry.
Thus, it is natural to ask how the pion properties would be modified in a nuclear medium,
where chiral symmetry is expected to be (partially) restored
(see Ref.~\cite{Kienle} for a review concerning this question).

There exist several works on pion properties in a nuclear medium, e.g. using the Nambu-Jona-Lasinio (NJL) model~\cite{NJL}, studies were made for
the dynamical mass (of the pion-like mode)~\cite{HatsudaKuni},
the pion structure function~\cite{Suzuki}, and the mass and
decay constant~\cite{Vogl,Hatsuda,Caldas}.
Other studies dealt with the pion decay constant in a composite-operator
approach~\cite{Barducci}, pion cloud effects on the Drell-Yan scattering~\cite{Dieperink}, mass shifts via in-medium chiral perturbation theory~\cite{Kirchbach,Meissner,Goda},
masses and decay constants within a Dyson-Schwinger and Bethe-Salpeter equation ansatz~\cite{Maris}.
Furthermore, the pion in finite density has been studied with QCD sum-rule approaches~\cite{Kim},
using virial expansions~\cite{Mallik}, a non-local chiral quark model~\cite{Nam},
a relativistic mean field approach~\cite{Roy}, and by exploiting operator relations
in QCD~\cite{Jido}. However, only one of them~\cite{Roy}
examined the in-medium modification of the pion
elastic form factor (in asymmetric nuclear matter) based on the $\rho-\omega$
mixing mechanism at the hadronic level. In the present work, we investigate
the pion structural properties in symmetric nuclear matter based on quark degrees of freedom.

In order to do so, it is necessary to have a reliable pion model which is well
constrained and successful in describing its properties in vacuum.
However, because of the Nambu-Goldstone boson nature of the pion with an anomalously
small mass, its description in vacuum itself is not straightforward either,
and a special treatment is necessary. Furthermore, even if such a successful model is available,
one requires a proper description of the system's state
as well as a consistent current operator to perform a meaningful comparison
with experimental data. To properly define the state of a relativistic system,
three space-time hyper-surfaces were identified by
Dirac~\cite{dirac}. These hyper-surfaces correspond to different forms of relativistic Hamiltonian dynamics, namely instant form, front form and point form.

In the present study, we apply the front-form dynamics incorporating consistency between the current
operator and the state from a field theoretical point of view with a fixed number of particles.
In particular, although the state has an infinite number of components in the Fock-space~\cite{brodsky},
only the lowest Fock component or valence component is usually modelled and used for
calculating electroweak form factors.
In principle, the infinite set of coupled eigenvalue equations for the Hamiltonian operator
in the Fock space can be replaced by an effective squared-mass operator or an effective Hamiltonian
acting on the valence sector. At the same time, it is also possible to express systematically the higher Fock-state
components of the wave function as functionals of the lower ones~\cite{brodsky,pauli3,pauli4}.
The effective electroweak current operator for the valence component of the state can also be consistently
derived within the field theoretical framework of the Bethe-Salpeter equation projected on the equal
light-front time~\cite{sales1,sales2}.
However, the major advance in the extraction of the form factors
from the front-form wave function is the realization that in the Drell-Yan frame $(q^+=0)$  the pair production
does not contribute unlike in the $q^+\ne 0$ frame~\cite{bakker01}.

In Ref.~\cite{Pacheco1}, the effect of such pair-terms was studied in detail
to evaluate the form factors of a system with two identical fermions,
where effective constituent quark masses of the lowest Fock-space
component were used to describe the pion properties in vacuum.
The vertex function describes the momentum component of the coupling
of the quarks to the pion bound state, where by momentum component
we mean the light-front wave function obtained by integrating over the relative
quark momentum, $k^-$, after the separation of the instantaneous terms in the
external quark legs.
Namely, the momentum component is the light-front
wave function which depends on the kinematic variables, $k^+$ and $\vec{k}_\perp$.
In this model, the symmetric vertex function and the light-front valence wave function
which are symmetric under the exchange of the fermion momenta are employed~\cite{Pacheco1}.
(See also Ref.~\cite{pipach} for a nonsymmetric case.) Symmetry properties of a Bethe-Salpeter
amplitude are derived from quantum field theory, where conceptual and phenomenological problems arise
when a nonsymmetric vertex function is used to describe the pion~\cite{pipach,bakker01}.
For example, the form factor and weak decay constant cannot be reproduced simultaneously.
In this study, we use the pion model of Ref.~\cite{Pacheco1}
which has all the desired properties discussed above.

The main purpose of the present article is to investigate the in-medium modifications
of the pion properties, i.e. the electromagnetic form factor, radius and decay constant
in symmetric nuclear matter, where the pion model~\cite{Pacheco1}
is adjusted so as to provide the best description of the electromagnetic form factor data in vacuum.
Because the (symmetric) nuclear matter is translational and rotational invariant,
usual Lorentz transformation can be applied. Thus, the input obtained
in the nuclear matter rest frame by the nuclear matter model
we employ, keeps the track of the Lorentz scalar and vector nature of mean fields,
and thus our assumption on the pion vertex model enables us to extract
the form factor in a covariant manner.

For the nuclear matter, we employ the quark-meson coupling (QMC) model~\cite{Guichon,QMCfinite,QMCreview}
in order to include consistently the in-medium properties of the constituent up and down quarks in the pion
in symmetric nuclear matter. The QMC model has been widely applied
to various nuclear and hadronic phenomena in a nuclear medium with successes.
Although our approach may be regarded as crude, this is a first exploratory study
that treats both the constituent quarks forming nuclear matter in the bound nucleons
and those in the pion immersed in this nuclear medium on the same footing.
In particular, it is known that a treatment of nuclear matter based on
quark degrees of freedom is very difficult and a description starting from
first principles of QCD is far remote.
In this sense, although phenomenological, this study may give some insight
in the pion properties in a nuclear medium.

This article is organized as follows. In Section  II we briefly describe the QMC model focusing on the properties of
constituent up and down quarks and the pion vertex in symmetric nuclear matter. The expressions for the in-medium
electromagnetic form factor of the pion are discussed in Section III, while the results for the in-medium pion properties,
electromagnetic form factor, radius and weak decay constant are presented in Section IV. Finally, Section V is devoted
to a summary and discussions.

\section{Quarks in nuclear matter}
\label{qmatter}

In order to study consistently the modifications of the pion properties in a nuclear medium, we need a reasonable
model of nuclear matter based on the quark degrees of freedom, as well as a pion model which successfully
describes the pion properties in vacuum. We first discuss the quark model description of nuclear matter
with the QMC model, with presenting some results calculated for symmetric nuclear matter.

\subsection{Quark Model of Nuclear Matter: QMC Model}
\label{QMC}

The QMC model was introduced in 1988 by Guichon~\cite{Guichon} with the MIT bag model,
and by Frederico {\it et al\/}. in 1989~\cite{Frederico} with help of a confining harmonic potential,
both approaches to describe nuclear matter properties based
on the quark degrees of freedom.
The model has been successfully applied for various studies of finite
(hyper)nuclei~\cite{QMCfinite} as well as the hadron properties in a nuclear medium
(see Ref.~\cite{QMCreview} for a comprehensive review).
In the model the medium effects arise through the self-consistent coupling of
phenomenological isoscalar-Lorentz-scalar ($\sigma$),
isoscalar-Lorentz-vector ($\omega$) and isovector-Lorentz-vector ($\rho$)
meson fields to the confined light-flavor $u$ and $d$ valence quarks --- rather than to the nucleons.
As a result the internal structure of the bound nucleon is modified by the
surrounding nuclear medium with respect to the free nucleon case.

The effective Lagrangian density for a uniform, spin-saturated,
and isospin-symmetric nuclear system (symmetric nuclear matter)
at the hadronic level is given by~\cite{Guichon,QMCfinite,QMCreview},
\begin{equation}
{\cal L} = {\bar \psi} [i\gamma \cdot \partial -m_N^*({\hat \sigma}) -g_\omega {\hat \omega}^\mu \gamma_\mu ] \psi
+ {\cal L}_\textrm{meson} ,
\label{lag1}
\end{equation}
where $\psi$, ${\hat \sigma}$ and ${\hat \omega}$ are respectively the nucleon,
Lorentz-scalar-isoscalar $\sigma$, and Lorentz-vector-isoscalar $\omega$ field operators, with
\begin{equation}
m_N^*({\hat \sigma}) = m_N - g_\sigma({\hat \sigma}) {\hat \sigma} ,
\label{efnmas}
\end{equation}
which defines the $\sigma$-field dependent coupling constant,
$g_\sigma({\hat \sigma})$, while $g_\omega$ is the nucleon-$\omega$ coupling
constant. All the important effective nuclear many-body dynamics including 3-body
nucleon force modeled at the quark level, will effectively be condensed in $g_\sigma({\hat \sigma})$.
Solving the Dirac equations for the up and down quarks in the nuclear medium with
the same mean fields (mean values) $\sigma$ and $\omega$, which act on the bound nucleon self-consistently
based on Eq.~(\ref{lag1}), we obtain the effective $\sigma$-dependent coupling $g_\sigma(\sigma)$
at the nucleon level~\cite{Guichon,QMCfinite,QMCreview}. The free meson Lagrangian density is given by,
\begin{equation}
{\cal L}_\mathrm{meson} = \frac{1}{2} (\partial_\mu {\hat \sigma} \partial^\mu {\hat \sigma} - m_\sigma^2 {\hat \sigma}^2)
- \frac{1}{2} \partial_\mu {\hat \omega}_\nu (\partial^\mu {\hat \omega}^\nu - \partial^\nu {\hat \omega}^\mu)
+ \frac{1}{2} m_\omega^2 {\hat \omega}^\mu {\hat \omega}_\mu \ ,
\label{mlag1}
\end{equation}
where we have ignored the isospin-dependent Lorentz-vector-isovector $\rho$-meson field,
since we consider isospin-symmetric nuclear matter within the Hartree mean-field approximation.
In this case the mean value of the $\rho$-mean field becomes zero and there is
no need to consider its possible contributions due to the $\rho$-Fock (exchange) terms.

In the following we work in the nuclear matter rest frame.
For symmetric nuclear matter in the mean-field approximation, the nucleon Fermi momentum $k_F$
(baryon density $\rho$) and the scalar density ($\rho_s$) associated with
the $\sigma$-mean field can be related as,
\begin{eqnarray}
\rho &=& \frac{4}{(2\pi)^3}\int d\vec{k}\ \theta (k_F - |\vec{k}|)
= \frac{2 k_F^3}{3\pi^2},
\label{rhoB}\\
\rho_s &=& \frac{4}{(2\pi)^3}\int d\vec{k} \ \theta (k_F - |\vec{k}|)
\frac{m_N^*(\sigma)}{\sqrt{m_N^{* 2}(\sigma)+\vec{k}^2}},
\label{rhos}
\end{eqnarray}
where $m_N^*(\sigma)$ is the constant value of the effective nucleon
mass at a given density and is calculated in the quark model.
In the standard QMC approach~\cite{Guichon,QMCfinite,QMCreview},
one uses the MIT bag model and the Dirac equations for the up and down
quarks in symmetric nuclear matter are solved
self-consistently with the $\sigma$ and $\omega$ mean-field potentials.
The Dirac equations for the quarks and antiquarks
($q=u$ or $d$, and $Q=s,c$ or $b$, quarks)
in the bag of hadron $h$ in nuclear matter at the position
$x=(t,\vec{r})$ ($|\vec{r}| \le$ bag radius) are given by~\cite{QMCreview},
\begin{eqnarray}
\left[ i \gamma \cdot \partial_x -
(m_q - V^q_\sigma)
\mp \gamma^0
\left( V^q_\omega +
\frac{1}{2} V^q_\rho
\right) \right]
\left( \begin{array}{c} \psi_u(x)  \\
\psi_{\bar{u}}(x) \\ \end{array} \right) &=& 0,
\label{diracu}\\
\left[ i \gamma \cdot \partial_x -
(m_q - V^q_\sigma)
\mp \gamma^0
\left( V^q_\omega -
\frac{1}{2} V^q_\rho
\right) \right]
\left( \begin{array}{c} \psi_d(x)  \\
\psi_{\bar{d}}(x) \\ \end{array} \right) &=& 0,
\label{diracd}\\
\left[ i \gamma \cdot \partial_x - m_{Q} \right]
\psi_{Q} (x)\,\, ({\rm or}\,\, \psi_{\Qbar}(x)) &=& 0,
\label{diracQ}
\end{eqnarray}
where we have neglected the Coulomb force as usual, since the nuclear matter
properties are due to the strong interaction, and we assume SU(2) symmetry for the light quarks,
$m_q = m_u = m_d$, and define $m^*_q \equiv m_q - V^q_\sigma = m^*_u = m^*_d$.
In symmetric nuclear matter, the isospin dependent $\rho$-meson mean field
in Hartree approximation yields $V^q_\rho = 0$ in Eqs.~(\ref{diracu})
and~(\ref{diracd}), so we ignore it hereafter.
The constant mean-field potentials in nuclear matter are defined by,
$V^q_\sigma \equiv g^q_\sigma \sigma = g^q_\sigma <\sigma>$ and
$V^q_\omega \equiv g^q_\omega \omega = g^q_\omega\, \delta^{\mu,0} <\omega^\mu>$,
with $g^q_\sigma$ and $g^q_\omega$ being the corresponding quark-meson coupling constants,
and the quantities inside the brackets stand for taking expectation values
by the nuclear matter ground state~\cite{QMCreview}.
Note that, since the velocity averages to zero in the rest frame of nuclear matter,
the mean vector source due to the quark fields as well,
$<\bar{\psi_q} \vec{\gamma} \psi_q> = 0$.
Thus we may just keep the term proportional to $\gamma^0$ in Eqs.~(\ref{diracu})
and~(\ref{diracd}).

The normalized, static solution for the ground state quarks or antiquarks
with flavor $f$ in the hadron $h$, may be written,
$\psi_f (x) = N_f e^{- i \epsilon_f t / R_h^*}
\psi_f (\vec{r})$,
where $N_f$ and $\psi_f(\vec{r})$
are the normalization factor and
corresponding spin and spatial part of the wave function.
The bag radius in medium for a hadron $h$, $R_h^*$,
is determined through the
stability condition for the mass of the hadron against the
variation of the bag radius~\cite{QMCreview}.
The eigenenergies in units of $1/R_h^*$ are given by,
\bge
\left( \begin{array}{c}
\epsilon_u \\
\epsilon_{\bar{u}}
\end{array} \right)
= \Omega_q^* \pm R_h^* \left(
V^q_\omega
+ \frac{1}{2} V^q_\rho \right),\,\,
\left( \begin{array}{c} \epsilon_d \\
\epsilon_{\bar{d}}
\end{array} \right)
= \Omega_q^* \pm R_h^* \left(
V^q_\omega
- \frac{1}{2} V^q_\rho \right),\,\,
\epsilon_{Q}
= \epsilon_{\Qbar} =
\Omega_{Q}.
\label{energy}
\ene

The hadron masses
in a nuclear medium $m^*_h$ (free mass $m_h$),
are calculated by
\begin{eqnarray}
m_h^* &=& \sum_{j=q,\bar{q},Q,\Qbar}
\frac{ n_j\Omega_j^* - z_h}{R_h^*}
+ {4\over 3}\pi R_h^{* 3} B,\quad
\left. \frac{\partial m_h^*}
{\partial R_h}\right|_{R_h = R_h^*} = 0,
\label{hmass}
\end{eqnarray}
where $\Omega_q^*=\Omega_{\bar{q}}^*
=[x_q^2 + (R_h^* m_q^*)^2]^{1/2}$, with
$m_q^*=m_q{-}g^q_\sigma \sigma$,
$\Omega_Q^*=\Omega_{\Qbar}^*=[x_Q^2 + (R_h^* m_Q)^2]^{1/2}$,
and $x_{q,Q}$ being the lowest bag eigenfrequencies.
$n_q (n_{\qbar})$ and $n_Q (n_{\Qbar})$
are the quark (antiquark)
numbers for the quark flavors $q$ and $Q$, respectively.
The MIT bag quantities, $z_h$, $B$, $x_{q,Q}$,
and $m_{q,Q}$ are the parameters for the sum of the c.m. and gluon
fluctuation effects, bag constant, lowest eigenvalues for the quarks
$q$ or $Q$, respectively, and the corresponding current quark masses.
$z_N$ and $B$ ($z_h$) are fixed by fitting the nucleon
(the hadron) mass in free space. (See table~\ref{Tab:QMC} the nucleon case.)

For the nucleon $h=N$
case in the above, the lowest, positive bag eigenfunction is given by
\begin{equation}
q(t,\vec{r}) = \frac{\cal N}{\sqrt{4\pi}}
e^{-i\epsilon_qt/R^*_N}\left( \begin{array}{c}
j_0(xr/R^*_N) \\
i\beta_q {\vec \sigma}\cdot{\hat r} j_1(xr/R^*_N)
\end{array} \right) \theta(R^*_N-r) \chi_m ,
\label{lowest}
\end{equation}
with $r=|\vec{r}|$ and $\chi_m$ the spin function and
\begin{eqnarray}
\Omega_q^* &=& \sqrt{x^2+(m_q^\ast R^*_N)^2}, \ \
\beta_q = \sqrt{\frac{\Omega_q^* - m_q^\ast R^*_N}{\Omega_q^* + m_q^\ast R^*_N}} ,
\label{omega0} \\
{\cal N}^{-2} &=& 2R_N^{*3} j_0^2(x)[\Omega_q^*(\Omega_q^*-1) + m_q^\ast R^*_N/2]/x^2  ,
\label{norm}
\end{eqnarray}
where $x$ is the eigenvalue for the lowest mode,
which satisfies the boundary condition at the bag surface,
$j_0(x) = \beta_q j_1(x)$.

The same meson mean fields $\sigma$ and $\omega$ for the quarks satisfy
the following equations at the nucleon level self-consistently:
\begin{eqnarray}
{\omega}&=&\frac{g_\omega \rho}{m_\omega^2},
\label{omgf}\\
{\sigma}&=&\frac{g_\sigma }{m_\sigma^2}C_N({\sigma})
\frac{4}{(2\pi)^3}\int d\vec{k} \ \theta (k_F - |\vec{k}|)
\frac{m_N^*(\sigma)}{\sqrt{m_N^{* 2}(\sigma)+\vec{k}^2}},
\label{sigf}\\
C_N(\sigma) &=& \frac{-1}{g_\sigma(\sigma=0)}
\left[ \frac{\partial m^*_N(\sigma)}{\partial\sigma} \right],
\label{CN}
\end{eqnarray}
where $C_N(\sigma)$ is the constant value of the scalar density ratio~\cite{Guichon,QMCfinite,QMCreview}.
Because of the underlying quark structure of the nucleon used to calculate
$M^*_N(\sigma)$ in the nuclear medium (see Eq.~(\ref{hmass}) with $h=N$),
$C_N(\sigma)$ gets $\sigma$-dependence,
whereas the usual point-like nucleon-based model yields unity, $C_N(\sigma) = 1$.
It is this $C_N(\sigma)$ or $g_\sigma (\sigma)$ that gives a novel saturation mechanism
in the QMC model, and contains the important dynamics which originates in the quark structure
of the nucleon. Without an explicit introduction of the nonlinear
couplings of the meson fields in the Lagrangian density at the nucleon and meson level,
the standard QMC model yields the nuclear incompressibility of $K \simeq 280$~MeV,
which is in contrast to a naive version of quantum hadrodynamics (QHD)~\cite{QHD}
(the point-like nucleon model of nuclear matter),
results in the much larger value, $K \simeq 500$~MeV;
the empirically extracted value falls in the range $K = 200 - 300$ MeV.
(See Ref.~\cite{Stone} for the updated discussions on the incompressibility.)

\begin{table}[t!]
\begin{center}
\caption{Coupling constants, the parameter $Z_N$, bag constant $B$ (in $B^{1/4}$),
and calculated properties for symmetric nuclear matter
at normal nuclear matter density $\rho_0 = 0.15$ fm$^{-3}$,
for $m_q = 5$ and $220$ MeV. The effective nucleon mass, $m_N^*$, and the nuclear
incompressibility, $K$, are quoted in MeV (the free nucleon bag radius used is $R_N = 0.8$ fm,
the standard value in the QMC model~\cite{QMCreview}). }
\label{Tab:QMC}
\bigskip
\begin{tabular}{c|cccccc}
\hline
$m_q$(MeV)&$g_{\sigma}^2/4\pi$&$g_{\omega}^2/4\pi$
&$m_N^*$ &$K$ & $Z_N$ & $B^{1/4}$(MeV)\\
\hline
 5   &5.39 &5.30 &754.6 &279.3 &3.295 &170 \\
 220 &6.40 &7.57 &698.6 &320.9 &4.327 &148 \\
\hline
\end{tabular}
\end{center}
\end{table}

Once the self-consistency equation for the ${\sigma}$, Eq.~(\ref{sigf}), has been solved, one can evaluate the total energy per nucleon:
\begin{equation}
E^\mathrm{tot}/A=\frac{4}{(2\pi)^3 \rho}\int d\vec{k} \
\theta (k_F - |\vec{k}|) \sqrt{m_N^{* 2}(\sigma)+
\vec{k}^2}+\frac{m_\sigma^2 {\sigma}^2}{2 \rho}+
\frac{g_\omega^2\rho}{2m_\omega^2} .
\label{toten}
\end{equation}
We then determine the coupling constants, $g_{\sigma}$ and $g_{\omega}$, so as
to fit the binding energy of 15.7~MeV at the saturation density $\rho_0$ = 0.15 fm$^{-3}$
($k_F^0$ = 1.305 fm$^{-1}$) for symmetric nuclear matter.

The pion model we adopt here~\cite{Pacheco1} uses a vacuum constituent quark mass, $m_q = 220$~MeV,
in order to well reproduce the electromagnetic form factor data and decay constant. Therefore, to be consistent
with this pion model, our nuclear matter is built with the same vacuum mass. The corresponding coupling
constants and some calculated properties for symmetric nuclear matter at the saturation density,
with the standard values of $m_{\sigma}=550$ MeV and $m_{\omega}=783$~MeV, are listed in Table~\ref{Tab:QMC}.
For comparison, we also give the corresponding quantities calculated in the standard QMC
model with a vacuum quark mass of $m_q = 5$~MeV (see Ref.~\cite{QMCreview} for details).
Thus we have obtained the necessary properties of the light-flavor constituent quarks in symmetric
nuclear matter with the empirically accepted data for a vacuum mass of $m_q = 220$~MeV; namely,
the density dependence of the effective mass (scalar potential) and vector potential.
The same in-medium constituent quark
properties will be used as input to describe the pion immersed in symmetric nuclear matter.

\begin{figure*}[t]
\begin{center}
\includegraphics[scale=0.6]{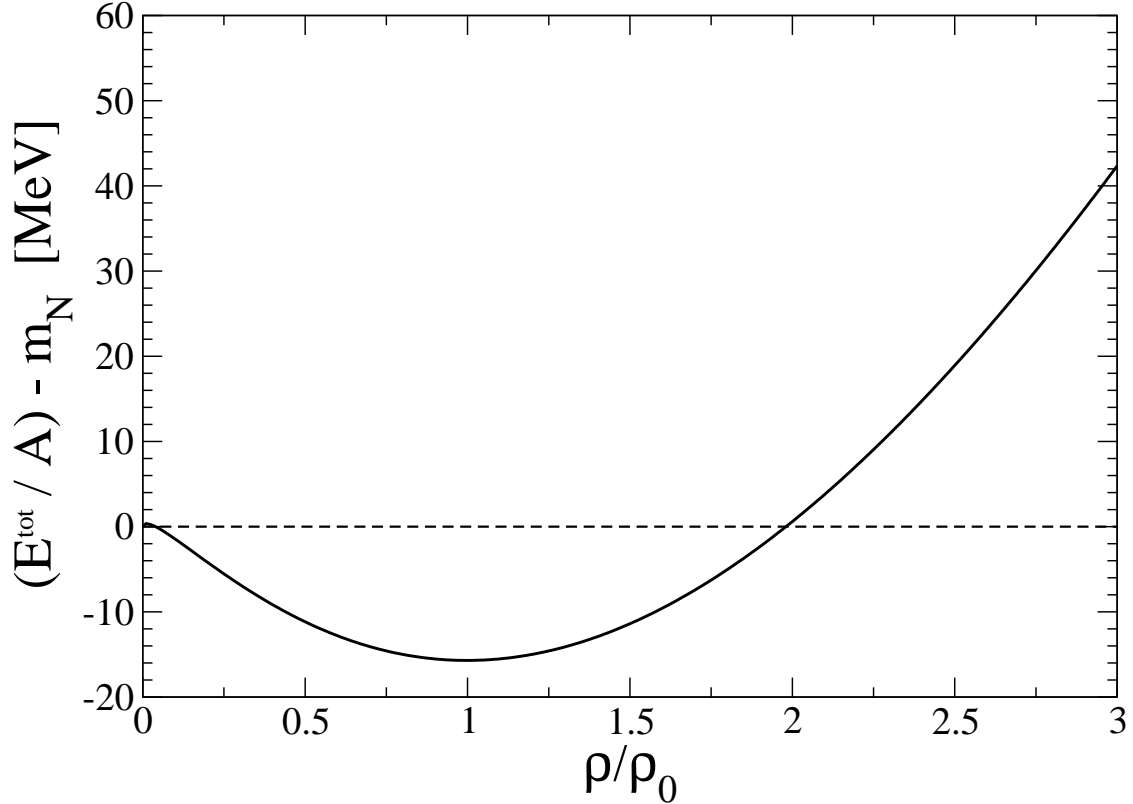}
\caption{Negative of the binding energy per nucleon ($E^\mathrm{tot}/A - m_N$) for symmetric
nuclear matter calculated with the vacuum up and down quark mass, $m_q = 220$ MeV.
At the saturation point $\rho_0 = 0.15$ fm$^{-3}$, the value is fitted to $-15.7$~MeV.
(See Ref.~\cite{QMCreview} for the $m_q = 5$ MeV case, denoted in there as QMC-I.)
\label{Fig:Eden}
}
\end{center}
\end{figure*}

\begin{figure*}[t]
\begin{center}
\includegraphics[scale=0.60]{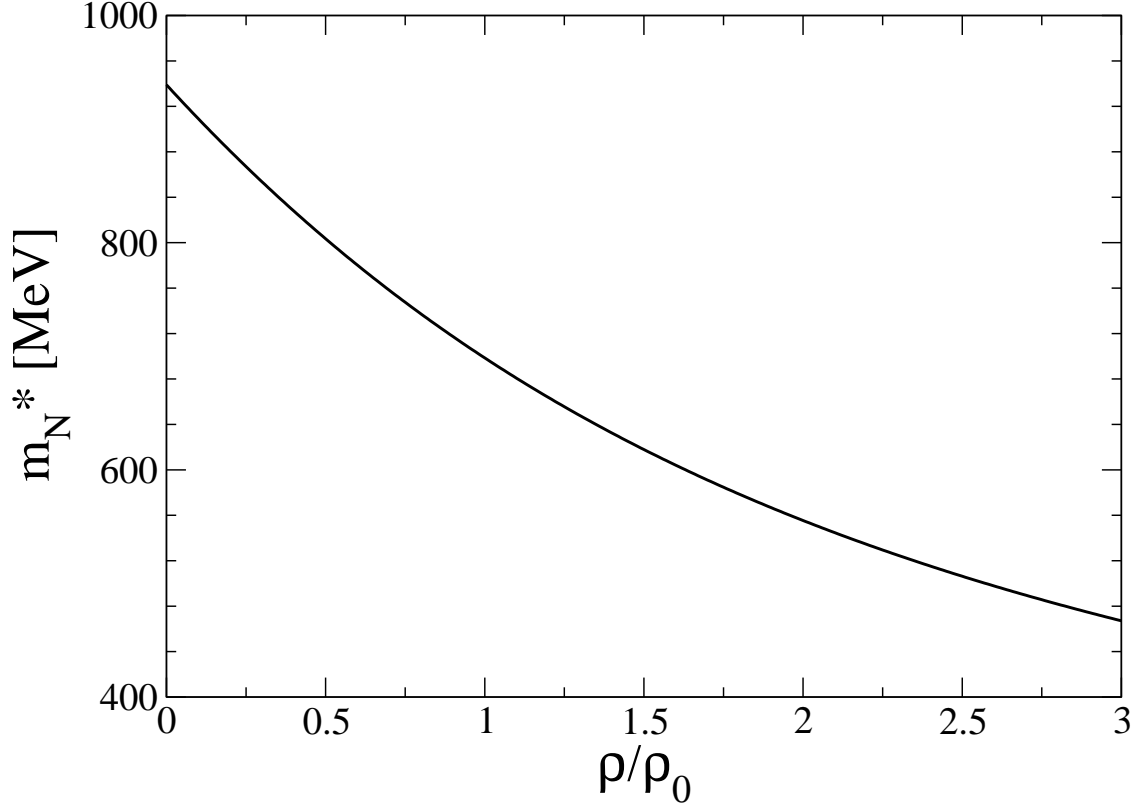}
\caption{Nucleon effective mass, $m^*_N$, in symmetric nuclear matter. See also caption of Fig.~\ref{Fig:Eden}.
\label{Fig:mNstar}
}
\end{center}
\end{figure*}

\begin{figure*}[t]
\includegraphics[scale=.60]{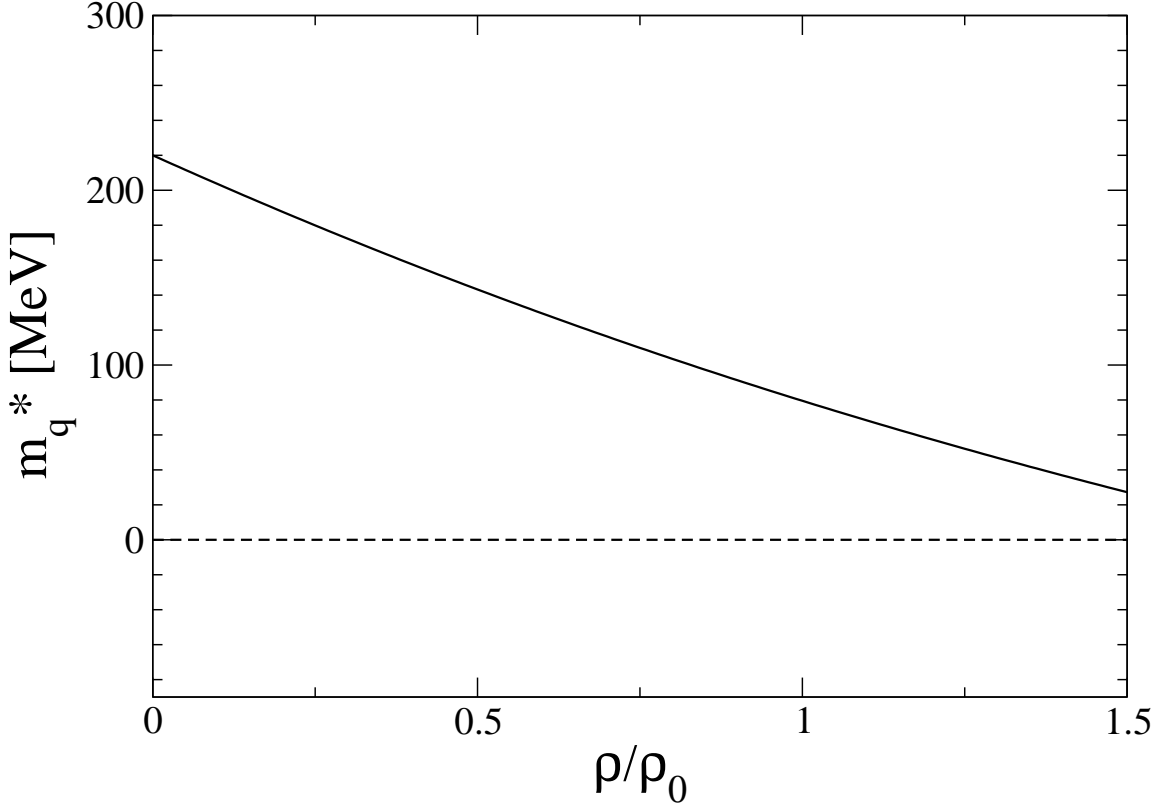}
\caption{Effective mass of the constituent up and down quarks, $m^*_q \equiv m^*_u = m^*_d$. See also caption of Fig.~\ref{Fig:Eden}.
}
\label{Fig:mqstar}
\end{figure*}

In Figs.~\ref{Fig:Eden},~\ref{Fig:mNstar} and~\ref{Fig:mqstar}, we respectively show our results for
the negative of the binding energy per nucleon ($E^\mathrm{tot}/A - m_N$),
effective mass of the nucleon, $m_N^*$,
and effective mass of the constituent up and down quarks, $m_q^*$, in symmetric nuclear matter.

%

As one can expect from the values of the incompressibility, $K = (279.3, 320.9)$~MeV in Table~\ref{Tab:QMC}, the result for
$E/A - m_N$ with $m_q = 220$ MeV shown in Fig.~\ref{Fig:Eden} varies slightly faster than that for
$m_q = 5$~MeV~\cite{QMCreview} with increasing density. As for the effective nucleon mass shown in
Fig.~\ref{Fig:mNstar} with $m_q = 220$~MeV, also decreases
faster than that for $m_q = 5$ MeV~\cite{QMCreview} with increasing nuclear density.

Concerning the effective constituent quark mass $m_q^*$ shown
in Fig.~\ref{Fig:mqstar}, a general comment in connection with the light-front model~\cite{Pacheco1,Pacheco2}
is in order: due to the pole structure of the propagators, the sum of the in-medium constituent
quark masses must be larger than the effective mass of the pion, $m^*_\pi$, namely
$2 m^*_q >  m^*_\pi$. Moreover, the pion mass up to normal nuclear matter density is
expected to be modified
only slightly, where the modification $\delta m_\pi$ at nuclear density
$\rho = 0.17$ fm$^{-3}$ averaged over the pion isospin states
is estimated as $\delta m_\pi \simeq +3$ MeV~\cite{Hayano,Kienle,Meissner,Vogl}.
Therefore, we approximate the effective
pion mass value to be the same as in vacuum,
$m^*_\pi \simeq m_\pi$ up to $\rho = \rho_0 = 0.15$ fm$^{-3}$.
In Fig.~\ref{Fig:mqstar} we present the results for the calculated effective mass
of the constituent quarks, $m^*_q$, up to $1.5 \rho_0$,
focusing on the relevant region where $2m^*_q > m^*_\pi \simeq m_\pi$ is satisfied.

\subsection{Quark Propagator and Pion Vertex in Symmetric Nuclear Matter}
\label{pionv}

In general, the quark self-energy in symmetric nuclear matter is modified by the Lorentz-scalar-isoscalar and Lorentz-vector-isoscalar
potentials. In the Hartree mean field approximation discussed in section~\ref{qmatter},
the modifications enter as
the shift of the quark (antiquark) momentum via
$p^\mu\to p^\mu + V^\mu = p^\mu + \delta^\mu_0V^0 \
(= p^\mu \pm \delta^\mu_0V^q_\omega$; $+(-)$ for quark(antiquark)) due to the vector potential, and in the Lorentz-scalar part through the
the Lorentz-scalar potential $V_s$ as $m_q \to m_q^* \equiv m_q + V_s \ (= m_q - V^q_\sigma)$. Since the Lorentz transformation properties
are retained in nuclear matter, these modifications can be implemented in the pion model~\cite{Pacheco1} without difficulties.
Then, the up or down quark propagator (containing the quark and antiquark components) in symmetric nuclear matter is given by,
\begin{eqnarray}
\displaystyle S^*(p+V)=\frac{1}{\rlap\slash p+\rlap\slash V-m_q^*+ i\epsilon},
\end{eqnarray}
while the in-medium pion vertex~\cite{Pacheco1} is modified as,
\begin{equation}
\Lambda^*(k+V,P)=
\frac{C^*}{((k+V)^2-m^2_{R} + i\epsilon)}+
\frac{C^*}{((P-k-V)^2-m^2_{R}+ i\epsilon)},
\label{vertex}
\end{equation}
where the normalization factor associated with $C^*$ is also modified by the medium effects. The regulator mass
$m_R$ represents soft effects at short range, namely at about the 1~GeV scale, and $m_R$ may also be influenced
by in-medium effects. However, since there exists no established way of estimating this effect
on the regulator mass and we already approximate $m_\pi^* = m_\pi$, we also employ $m_R^* = m_R$.
In addition, since it is correlated with the in-medium modified constant $C^*$ discussed in Section~\ref{mediumFF},
we use the vacuum regulator mass value $m_R$ to avoid introducing another source of uncertainty.

\section{The in-medium electromagnetic form factor model}
\label{mediumFF}

The electromagnetic interaction of  a pion, a spin-zero $q\overline{q}$ bound composite
system in vacuum, is based on three ingredients~\cite{Pacheco1};
{\it i\/}) effective Lagrangian which models the coupling of the pion
field to the quark fields, {\it ii\/}) a symmetric vertex function in momentum space,
{\it iii\/}) effective constituent quark masses and the lowest Fock state.
We follow the procedure in vacuum {\it i\/}) $-$ {\em iii\/}) with the in-medium
constituent quark properties as input and calculate the in-medium pion properties
using an effective Lagrangian density with a pseudoscalar coupling~\cite{tob92},
\begin{equation}
{{\cal L}_I= - i g^* \vec\Phi \cdot \overline q \gamma^5 \vec \tau q \, \Lambda^*,}  \label{lain}
\end{equation}
where $g^*$ is the coupling constant and $\Lambda^*$ is the in-medium vertex function.
The coupling constant $g^*$ is given by the Goldberger-Treiman relation at the quark level,
$g^* = m_q^*/f^*_\pi$, with the in-medium pion decay constant $f^*_\pi$.
The constant $C^*$ in Eq.~(\ref{vertex}) is determined from the charge normalization
for the spin-zero composite system and
it is density dependent. The photon field is coupled the minimal way satisfying current conservation.
The front-form coordinates
are defined as, $k^+ = (k+V)^0 + k^3,\: k^- = (k+V)^0 - k^3$,
and $\vec k_\perp\equiv(k^1,k^2).$

The electromagnetic current associated with the $\pi^+$ is obtained from the
corresponding Feynman triangle diagram;
\begin{equation}
\hspace*{-1mm}
j^\mu = -i\, 2 e \frac{m_q^{*2}}{f^{*2}_\pi} N_c\! \int\! \frac{d^4k'}{(2\pi)^4}\, \mathrm{Tr} \left [S^*(k')
\gamma^5 S^*(k'-P^{\prime}) \gamma^\mu S^*(k'-P) \gamma^5 \right ]\Lambda^*(k',P^{\prime}) \Lambda^*(k',P)\, ,
\label{jmu}
\end{equation}
where $(k')^\mu = k^\mu + \delta^\mu_0V^0$, and $N_c = 3$ is the number of colors. The factor 2 stems from isospin algebra.
(It is easy to prove that the Ward identity is satisfied in the Breit-frame: first one performs
the trace in $q\cdot j$, and notices that the integrand of the resulting expression
is odd in $\vec{k}' = \vec{k} \rightarrow -\vec{k}' = -\vec{k}$, and therefore $q\cdot j = 0$.)

\begin{figure*}[t]
\begin{center}
\includegraphics[scale=1.2]{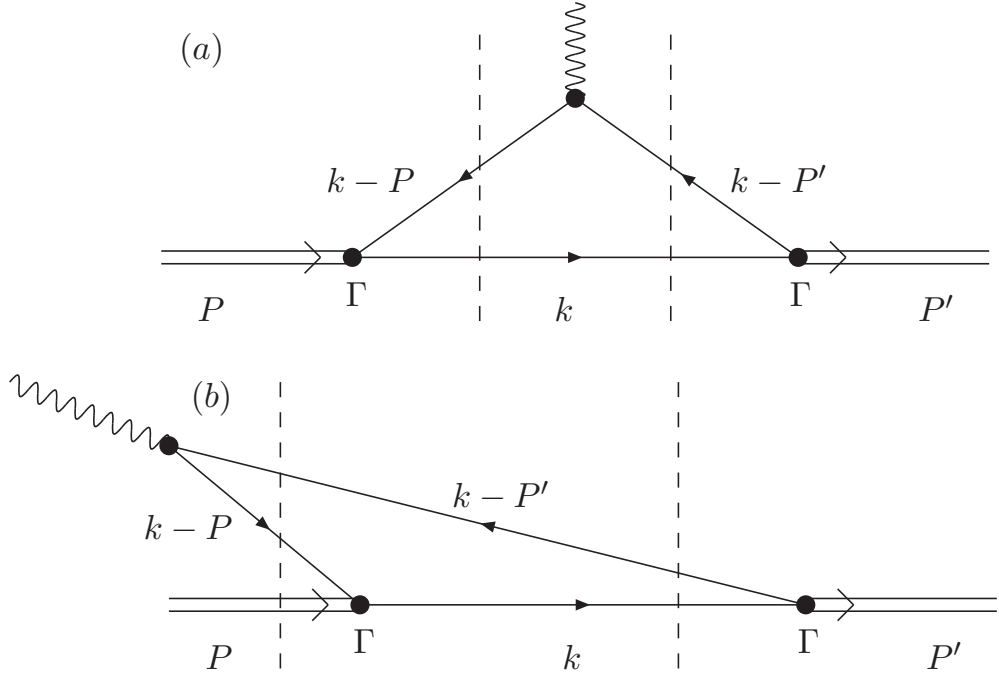}
\caption{Light-front time-ordered processes for the electromagnetic interaction of  the pion,
where the diagrams (a)  and (b) correpsond to $F_\pi^{*(I)}$ in Eq.~(\ref{FI})) and $F_\pi^{*(II)}$ in
Eq.~(\ref{FII}), respectively. The shift of variables, $k^\mu + V^\mu \to k^\mu$, is applied to
the loop integral.
\label{Fig:EM_Int}
}
\end{center}
\end{figure*}

We choose the symmetric vertex function, Eq.~(\ref{vertex}), which also produces a symmetric light-front
wave function under the interchange of the quark and antiquark momenta,
which improves the physical description without
the conceptual difficulties associated with the non-symmetric regulator (see also Refs.~\cite{bakker01,pach97}).
The normalization constant $C^*$ in Eq.~(\ref{vertex}) is fixed by imposing the condition $F^*_\pi(0)=1$ on the pion form factor.

The effect of the vector potential $\delta^\mu_0V^0$ in the loop integral cancels
identically due to the choice of the pion vertex. Therefore, only the mass shift of
the quarks is relevant in the  loop integral. In the four-momentum integration,
we apply the transformation, $k^{\prime\mu} = k^\mu + \delta^\mu_0V^0 \to k^\mu$,
so that no trace of the vector potential remains in the expressions of the
pion decay constant and electromagnetic form factor.
It is worthwhile to observe that, the current Eq.~(\ref{jmu}) means to be covariant,
while the quark propagator and pion vertex are computed in the rest frame of nuclear matter.
However, both the shift of the time component of the quark momentum and effective mass in the medium
allow us to recover a boost invariant form for the form factor as the shift in the virtual quark
energy can be absorbed by a variable change in the loop momentum, since we have assumed
that the vertex parameter is unchanged by the medium effects,
which may be justified by the fact that it corresponds to a short-range scale
deep inside the pion wave function, which is much smaller than the nucleon size.

We work in the Breit-frame, where the momentum transfer, $q^\mu=(P'-P)^\mu$, $q^2 = q^+q^--(\vec q_\perp)^2 \equiv -Q^2$, lies in the $z-x$ plane with $q^+=-q^-=\sqrt{-q^2}\sin \alpha$,  $q_x=\sqrt{-q^2}\cos \alpha$
and $q_y=0$ (the angle $\theta$ in Ref.~\cite{bakker01} corresponds to $\alpha+90^\circ$).
The initial and final momenta of the composite spin zero bound state with
mass $m_B$ are: $P^0=E=E^\prime=\sqrt{m^2_B -q^2/4}$, $\vec P^{\prime}_\perp=-\vec P_\perp= \frac{\vec q_\perp}{2}$ and $P^{\prime}_z=-P_z=\frac{q^+}{2}$.
The Drell-Yan condition $q^+=0$ is recovered with $\alpha=0^\circ$, while the $q^+=\sqrt{-q^2}$
condition~\cite{lev98} in the Breit-frame follows from $\alpha=90^{\circ}$. We here use $\alpha = 0^\circ$.

In general, the pion form factor in medium can be extracted from the covariant expression:
\begin{equation}
   j^\mu = e (P^{\mu}+P^{\prime \mu}) F^*_\pi (q^2).
\label{full}
\end{equation}
If covariance and current conservation are satisfied in the calculation, one can obviously compute the electromagnetic form factor in any frame and from any nonvanishing component of the current.

In the following, to compute the pion elastic form factor,
we use the pseudo-scalar Lagrangian density of Eq.~(\ref{lain}),
the current defined by Eq.~(\ref{jmu}), and the symmetric vertex function given
by Eq.~(\ref{vertex}) with the plus component
of the current, $j^+$, associated with Eq.~(\ref{jmu})
in the Breit-frame (with $\vec q$ in the $z-x$ plane).
Only two nonvanishing contributions in Eq.~(\ref{jmu}) contribute to the form factor~\cite{sawicki,pipach,bakker01,ji1}:
\begin{eqnarray}
F^*_\pi(q^2)=F^{*(I)}_\pi(q^2)+F^{*(II)}_\pi(q^2).
\label{ffactor}
\end{eqnarray}

With the replacement $k^\mu + \delta^\mu_0V^0 \to k^\mu$, the two contributions to the form factor obtained from $j^+$,
which correspond to the two diagrams shown in Fig.~\ref{Fig:EM_Int}, are given by the following expressions:
\begin{eqnarray}
F_\pi^{*(I)}(q^2)&=& - i
\frac{m_q^{*2}}{(P^{+}+P^{\prime +})f^{*2}_\pi}
\frac{N_c}{(2\pi)^4}
 \int \frac{d^{2} k_{\perp} d k^{+} d k^-\theta (k^+)\theta(P^+-k^+)}
{k^+(P^+-k^+) (P^{^{\prime}+}-k^+)}
\Pi^*(k,P,P'),
\label{FI}
\end{eqnarray}
and
\begin{eqnarray}
F_\pi^{*(II)}(q^2)&=& - i
\frac{m_q^{*2} }{(P^{+}+P^{\prime +})f^{*2}_\pi}
\frac{N_c}{(2\pi)^4}
 \int \frac{d^2 k_{\perp} d k^{+} d k^-\theta (k^+-P^+)
\theta (P^{\prime +}-k^+)}
{k^+(P^+-k^+) (P^{^{\prime}+}-k^+)} \Pi^*(k,P,P')\ ,
\nonumber\\
\label{FII}
\end{eqnarray}
where
\begin{eqnarray}
\Pi^* (k,P,P')&=& \frac{\mathrm{Tr}  [{\cal O^*}^+]\Lambda^*(k,P) \Lambda^*(k,P^\prime)}
{ (k^- - k^-_\mathrm{on}+ i\epsilon) (P^- - k^- -(P-k)^-_\mathrm{on}+ \frac{i\epsilon}{P^+ - k^+})}
\nonumber \\ &\times &
\frac{1}{(P^{\prime -} - k^- -(P^\prime-k)^-_\mathrm{on}+ i\epsilon)},
\label{pi}
\end{eqnarray}
with the ``in-medium on-the-energy shell''  values of the individual momentum given by
\begin{eqnarray}
k^-_\mathrm{on}=\frac{k_{\perp}^{2}+m_q^{*2}}{k^+} \ , \
(P-k)^-_\mathrm{on}=\frac{(P-k)_{\perp}^{2}+m_q^{*2}}{P^+-k^+}, \ \text{and} \
 (P^\prime-k)^-_\mathrm{on}=\frac{(P^\prime-k)_{\perp}^{2}+m_q^{*2}}{P^{\prime +}-k^+}.
\label{onek}
\end{eqnarray}
For the trace $\mathrm{Tr} [{\cal O}^{*+} ]$ of the operator in Eq.~(\ref{pi}),
\begin{eqnarray}
  {\cal O}^{*+}= (\rlap\slash k +m_q^*)\gamma^5(\rlap\slash k  - \rlap\slash P^\prime +m_q^*)
\gamma^+ (\rlap\slash k - \rlap\slash P + m_q^*) \gamma^5,
\label{O}
\end{eqnarray}
one finds,
\begin{eqnarray}
\tfrac{1}{4}\, \mathrm{Tr} [{\cal O}^{*+} ]  &=& -k^- (P^{\prime +}-k^+)(P^{+}-{k^+}) +(k^2_\perp+m_q^{*2})(k^+-P^+-P^{\prime +})
\nonumber \\
&-&\frac12 \vec k_\perp \cdot
(\vec P^{\prime }_\perp-\vec P_\perp) (P^{\prime +}-P^+)+\frac14 k^+q^2_\perp.
\label{tr}
\end{eqnarray}
The detailed forms of $F^{*(I)}$ and $F^{*(II)}$ in vacuum after integration over $k^-$ can be found in Appendices A and B of Ref.~\cite{Pacheco1}.

The explicit form of the symmetric regulator function in the front-form momentum coordinates in Eq.~(\ref{pi}) which enters in
Eqs.~(\ref{FI}) and~(\ref{FII})
is given by:
\begin{eqnarray}
\Lambda^* (k,P)&=&C^*\left[k^+ \left(k^--\frac{k^2_\perp+m^2_R- i\epsilon}{k^+} \right)
\right]^{-1}
\nonumber \\
&+&
C^*\left[(P^+-k^+) \left(P^--k^--\frac{(P-k)^2_\perp+m^2_R- i\epsilon}
{P^+-k^+}\right)\right]^{-1}.
\label{llf}
\end{eqnarray}

The sum of the contributions $F^{*(I)}_\pi$ and $F^{*(II)}_\pi$
in vacuum was already shown to yield the covariant result~\cite{Pacheco1}.
The different directions of $\vec q$ in the Breit-frame can only change the absolute values of $F_\pi^{*(I)}(q^2)$
and $F_\pi^{*(II)}(q^2)$, but not the sum. For example, with $q^+=0$ $(\alpha=0^\circ)$ we have
$F_\pi^{*(II)}(q^2)=0$, and thus $F_\pi^{*(I)}(q^2)$ alone yields the covariant result~\cite{Pacheco1}.

The in-medium quark Dirac propagator after the variable shift, $k^\mu + \delta^\mu_0V^0 \to k^\mu$, can be
decomposed using the front-form momenta~\cite{brodsky},
\begin{eqnarray}
\frac{\rlap\slash{k}+m_q^*}{k^2-m_q^{*2}+ i\epsilon}=
\frac{\rlap\slash{k}_\mathrm{on}+m_q^*}{k^+(k^--k^-_\mathrm{on}+\frac{i\epsilon}{k^+})} +\frac{\gamma^+}{2k^+},
\label{inst}
\end{eqnarray}
where $k^-_\mathrm{on}=(k_\perp^2+m_q^{*2})/k^+$. The second term on the right-hand side of Eq.~(\ref{inst}) is an
instantaneous term in the light-front time. The instantaneous term contributes to both, $F_\pi^{*(I)}(q^2)$ and $F_\pi^{*(II)}(q^2)$,
due to the analytic structure of the symmetric vertex function of Eq.~(\ref{vertex}). These contributions are of nonvalence nature,
as they are not reducible to the impulse approximation within the valence wave function.

\section{Valence Light-front wave function}
\label{LFwf}

The valence component of the light-front wave function in vacuum was obtained
in Ref.~\cite{Pacheco1}. The external two-fermion space-time coordinates of
the Bethe-Salpeter amplitude are constrained to equal light-front time
after dropping the instantaneous terms of the external Dirac propagators~\cite{sales2}.
However, the effect of the instantaneous terms in a Bethe-Salpeter
approach is included in the effective operators, together with the valence wave function~\cite{sales2}.
In the present treatment, the Bethe-Salpeter amplitude with the in-medium pion vertex of Eq.~(\ref{vertex})
can be written as~\cite{Pacheco1},
\begin{eqnarray}
\Psi ^*(k+V,P) = \frac{\rlap\slash{k}+\rlap\slash V+m_q^*}{(k+V)^2-m_q^{*2}+ i\epsilon}
\gamma^5 \Lambda^* (k+V,P)
\frac{\rlap\slash{k}+\rlap\slash V-\rlap\slash{P}+m_q^*}{(k+V-P)^2-m_q^{*2}+ i\epsilon}.
\label{bsa}
\end{eqnarray}
Separating out the instantaneous terms in the quark propagators as well as the remaining spinor
operator part in the numerator of Eq.~(\ref{bsa}) and the $k^+$ and $(P^+-k^+)$ factors
in Eq.~(\ref{bsa}), the momentum part (the part depends on the plus and transverse momenta)
of the valence component of the light-front wave function
with $k^\mu + \delta^\mu_0V^0 \to k^\mu$ is given by,
\begin{eqnarray}
\Phi^*(k^+,\vec k_\perp; P^+,\vec P_\perp)  &=&  i N^*   \int \frac{dk^-}{2\pi}
\frac{1}{(k^--k^-_\mathrm{on}+ \frac{i\epsilon}{k^+}) (P^--k^--(P-k)^-_\mathrm{on}+\frac{i\epsilon}{P^+-k^+})}
\nonumber \\
& \times & \left( \frac{1}{k^2-m_R^2+ i\epsilon} +\frac{1}{(P-k)^2-m_R^2+ i\epsilon} \right),
\label{wf1}
\end{eqnarray}
where $N^*$ is a normalization factor,
$$N^*=C^*\frac{m_q^*}{f^*_\pi}(N_c)^\frac12.$$
Performing the $k^-$ integration in Eq.~(\ref{wf1}), one has
\begin{eqnarray}
\Phi^*(k^+,\vec k_\perp; P^+,\vec P_\perp)=
\frac{P^+}{m^{*2}_\pi-M^2_0} & &  
~\left[\frac{N^*}
{(1-x)(m^{*2}_{\pi}-{\cal M}^2(m_q^{*2}, m_R^2))} \right.
\nonumber \\
& & \hspace{3em} \left. +\frac{N^*}
{x(m^{*2}_{\pi}-{\cal M}^2(m^{2}_R, m_q^{*2}))} \right],
\label{wf2}
\end{eqnarray}
where $x=k^+/P^+$, with \ $0 \le x \le 1$; 
${\cal M}^2(m^2_a, m_b^2)= \frac{k^2_\perp+m_a^2}{x}+\frac{%
(P-k)^2_\perp+m^2_{b}}{1-x}-P^2_\perp \ $,
and the square of the mass is $M^2_0 ={\cal M}^2(m_q^{*2}, m_q^{*2})$. 
Since the momentum part of the wave function
is symmetric under the exchange of the fermion momenta, we have a second term in Eq.~(\ref{wf2}),
which is different from Ref.~\cite{pipach}.

Using only the valence component, the electromagnetic form factor evaluated in the Breit-frame reads~\cite{pipach,tob92},
\begin{eqnarray}
F_\pi^{*(WF)}(q^2)= \frac{1}{2\pi^3(P^{\prime +}+P^+)}
        & &  \hspace{-1.5em} \int \frac{d^{2} k_{\perp} d k^{+} \theta (k^+)\theta(P^+-k^+)}{k^+(P^+-k^+) (P^{^{\prime}+}-k^+)}
         \Phi^*(k^+,\vec k_\perp;P^{\prime +},\tfrac{\vec q_\perp}{2})  \nonumber \\
&\times& \left (k^-_\mathrm{on}P^+P^{\prime +}-\tfrac{1}{2} \vec k_\perp \cdot \vec q_\perp (P^+-P^{\prime +})-\tfrac{1}{4}\, k^+q^2_\perp \right  )
\nonumber \\
&\times& \Phi^*(k^+,\vec k_\perp;P^{ +},-\tfrac{\vec q_\perp}{2}) \ .
\label{Fwf}
\end{eqnarray}
Once the normalization constant $C^*$ is obtained from the condition $F^*_\pi(0)=1$ (see Eq.~(\ref{ffactor})), the probability of
the valence $q\overline{q}$ component for the pion in medium can be calculated by setting $\eta^*=F_\pi^{*(WF)}(0)$.

For convenience, we introduce the transverse momentum probability density,
\begin{eqnarray}
f^*(k_\perp)= \frac{1}{4\pi^3 m^*_\pi} \int_0^{2\pi} d\phi \int^{P^+}_0 \frac{d k^{+}M_0^{*2}}
{k^+(P^+-k^+)} \Phi^{*2}(k^+,\vec k_\perp;m^*_\pi,\vec 0),
\label{prob1}
\end{eqnarray}
and  integration of $f^*(k_\perp)$ leads to the in-medium probability of the valence component in the pion:
\begin{eqnarray}
\eta^*=\int^\infty_0 dk_\perp k_\perp f^*(k_\perp).
\label{Eq:vcompo}
\end{eqnarray}
The in-medium pion decay constant, $f^*_\pi$, is defined as the matrix element of the partially conserved axial vector current
in symmetric nuclear matter, with the  ground state $|0(\rho)\rangle$:
\begin{equation}
P_\mu \langle 0(\rho)|A^\mu_i |\pi^*_j \rangle  = i m_\pi^{*2} f^*_\pi \delta_{ij}
\simeq i m_\pi^2 f^*_\pi \delta_{ij}.
\end{equation}
Using $A^\mu_i = \bar{q} \gamma^\mu \gamma^5 \frac{\tau_i}{2} q$ and the interaction Lagrangian
density, Eq.~(\ref{lain}), for the pion-$q\overline{q}$ vertex function,
we obtain after integration over $k^-$ the in-medium decay constant, $f^*_\pi$, in terms of
the valence component of the model~\cite{tob92},
\begin{eqnarray}
f^*_{\pi} = \frac{m_q^*(N_c)^\frac12}{4\pi^3} \int \frac{d^{2} k_{\perp} d k^+ } {k^+(P^+-k^+)}\,
\Phi^*(k^+,\vec k_\perp;m^*_\pi,\vec 0),
\label{fpi}
\end{eqnarray}
where $f^*_\pi$ above is associated with the plus-component on the light-front, i.e.
the light-front time component,  thus
the $f^*_\pi$ cannot be separated into time and space components as done in chiral perturbation
theory~\cite{Hayano,Kienle,Vogl,Kirchbach,Meissner,Goda}. The normalization condition of $\Phi^*$ is given by
the probability of finding the pion in the valence component state, $\eta^* = F_\pi^{*(WF)}(0)$, which is less than one,
similarly to the vacuum case~\cite{Pacheco1}. However, an interesting feature due to the in-medium effect arises,
which will be discussed in Section~\ref{Nresults}.

\section{Numerical Results}
\label{Nresults}

The pion model in vacuum has two free parameters, the constituent quark mass, $m_q=220$~MeV used
in meson phenomenology~\cite{tob92,salme,gi}, and the regulator mass, $m_R=600$~MeV obtained from
fitting Eq.~(\ref{fpi}) to the experimental value of $f^\mathrm{exp}_\pi=92.4$~ MeV~\cite{PDG}.
(In fact, the model yields $f_\pi = 93.1$ MeV with these parameter values,
whereas to reproduce exactly the value $f^\mathrm{exp}_\pi=92.4$~MeV some fine-tuning is necessary).
Recall that we approximate  the in-medium pion mass, $m^*_\pi \simeq m_\pi=140$~MeV, based on the
analyses of Refs.~\cite{Hayano,Vogl,Meissner} and empirical extraction~\cite{Kienle} from  pionic-atom data.

The squared-charge radius of the pion is derived from the elastic form factor,
\begin{equation}
\langle r_\pi^2 \rangle= \left. - 6\frac{\partial}{\partial q^2} F_\pi (q^2) \right|_{q^2 \to 0},
\end{equation}
and one obtains $\langle r_\pi^2 \rangle^{1/2} = 0.74$ fm in vacuum~\cite{Pacheco1}, to be compared
with the experimental value $0.67\pm 0.02$ fm~\cite{amen}. In practice, since the derivative is evaluated
numerically and $|F_\pi (q^2)|$ varies quite rapidly near $q^2 = 0$ for all chosen nuclear densities as well as
in vacuum, the values quoted in this work are all evaluated at $Q^2 = -q^2 = 0.001$~(GeV/$c$)$^2$,
where the stability of the form factor has been checked.

\begin{figure*}[t]
\begin{center}
\includegraphics[scale=.60]{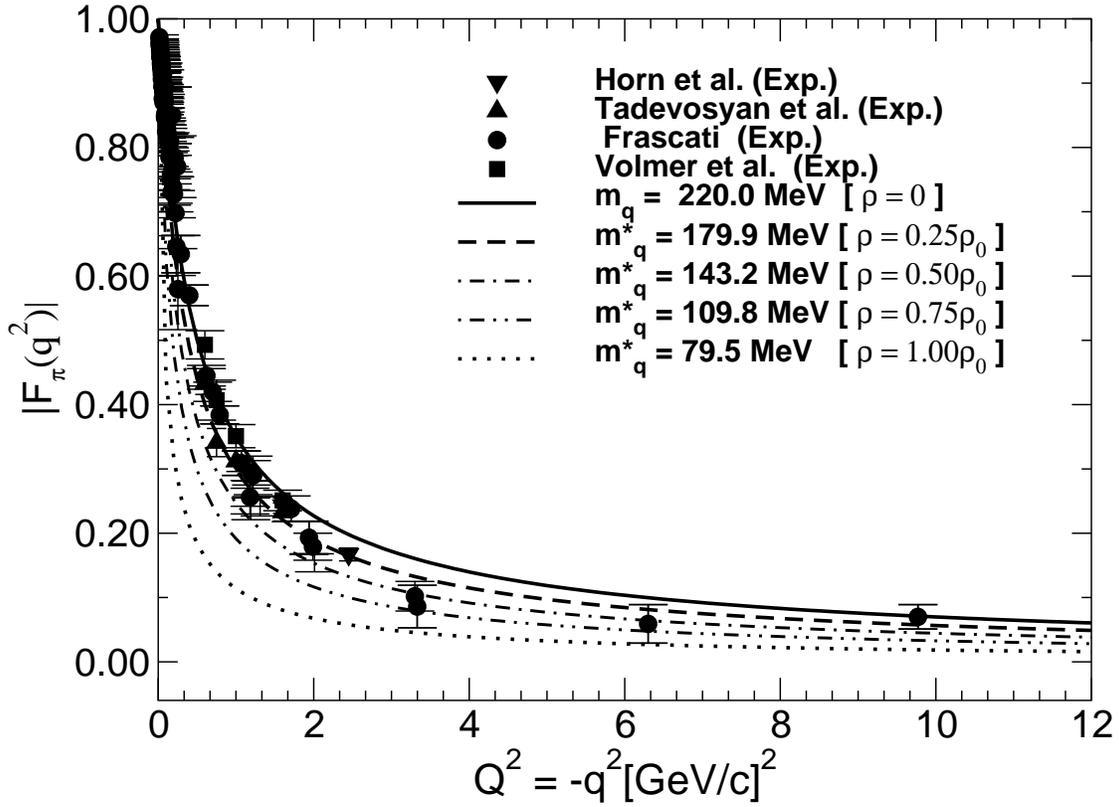}
\caption{Pion electromagnetic form factor in symmetric nuclear matter for four nuclear densities and the vacuum case as a function of  $Q^2=-q^2$. Experimental data in vacuum are from
Refs.~\cite{tj,cea,corn1,corn2,bebek}. Also shown are the
corresponding effective quark mass values, $m^*_q$, where in vacuum $m_q = 220$~MeV.}
\label{Fig:FF}
\end{center}
\end{figure*}

\begin{figure*}[t]
\begin{center}
\includegraphics[scale=.60]{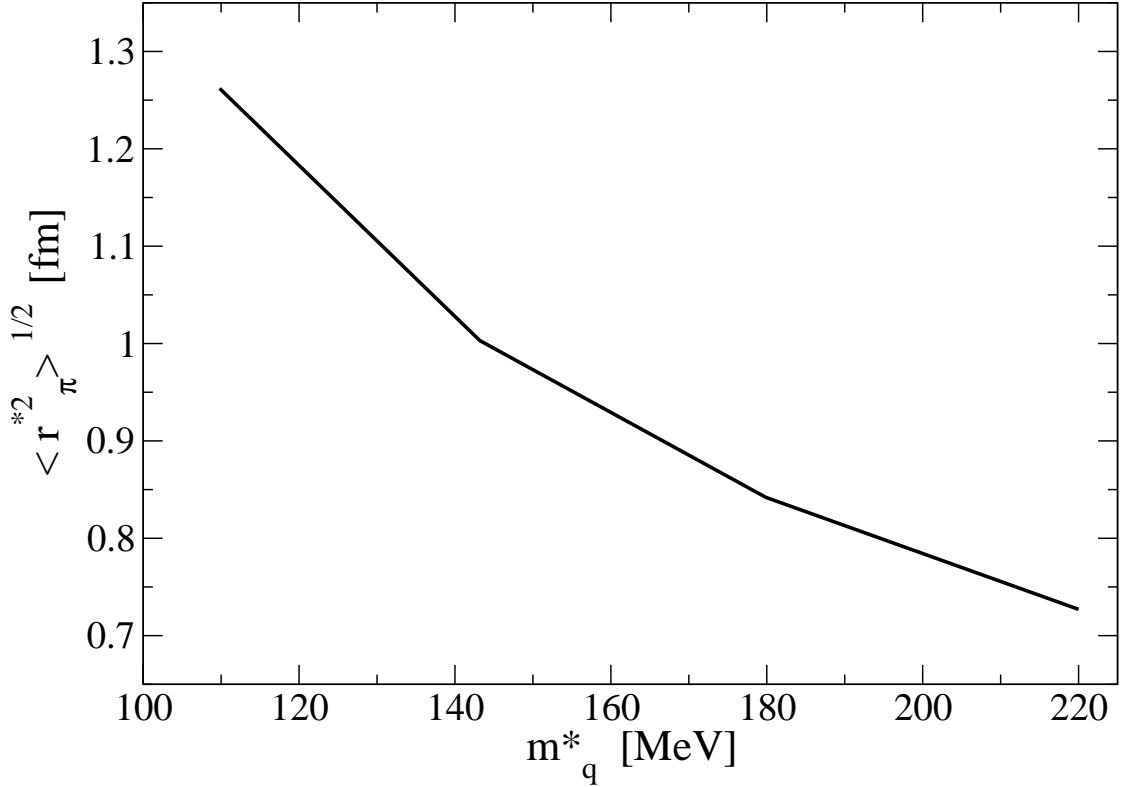}
\caption{The $m^*_q$ dependence of the pion charge radius, $\langle r^{*2}_\pi \rangle^{1/2}$.}
\label{Fig:rpimq}
\end{center}
\end{figure*}

In Fig.~\ref{Fig:FF}, the $Q^2$-dependence of the elastic form factor calculated in symmetric nuclear matter for
four nuclear densities along with  the vacuum case is presented. The experimental data in vacuum are from
Refs.~\cite{tj,cea,corn1,corn2,bebek} and the vacuum result agrees well with the data points of Ref.~\cite{tj}.
As the nuclear density increases, the absolute value of the form factor $|F_\pi(q^2)|$ becomes {\em harder}.
This  leads to a larger pion charge radius in nuclear matter with increasing density.
In Fig.~\ref{Fig:rpimq}, we show the $m^*_q$ dependence of the in-medium pion charge
radius, $\langle r^{*2}_\pi \rangle^{1/2}$.

From Figs.~\ref{Fig:FF} and~\ref{Fig:rpimq} one can see that the pion charge radius grows
as the nuclear density (effective quark mass) increases (decreases). The decrease in the constituent quark mass
kinematically allows for the quarks to move in a larger space region and the quark-antiquark bound
state becomes shallower; i.e. the pion is less bound which results in an increase of the charge radius.

\begin{figure*}[t]
\begin{center}
\includegraphics[scale=.60]{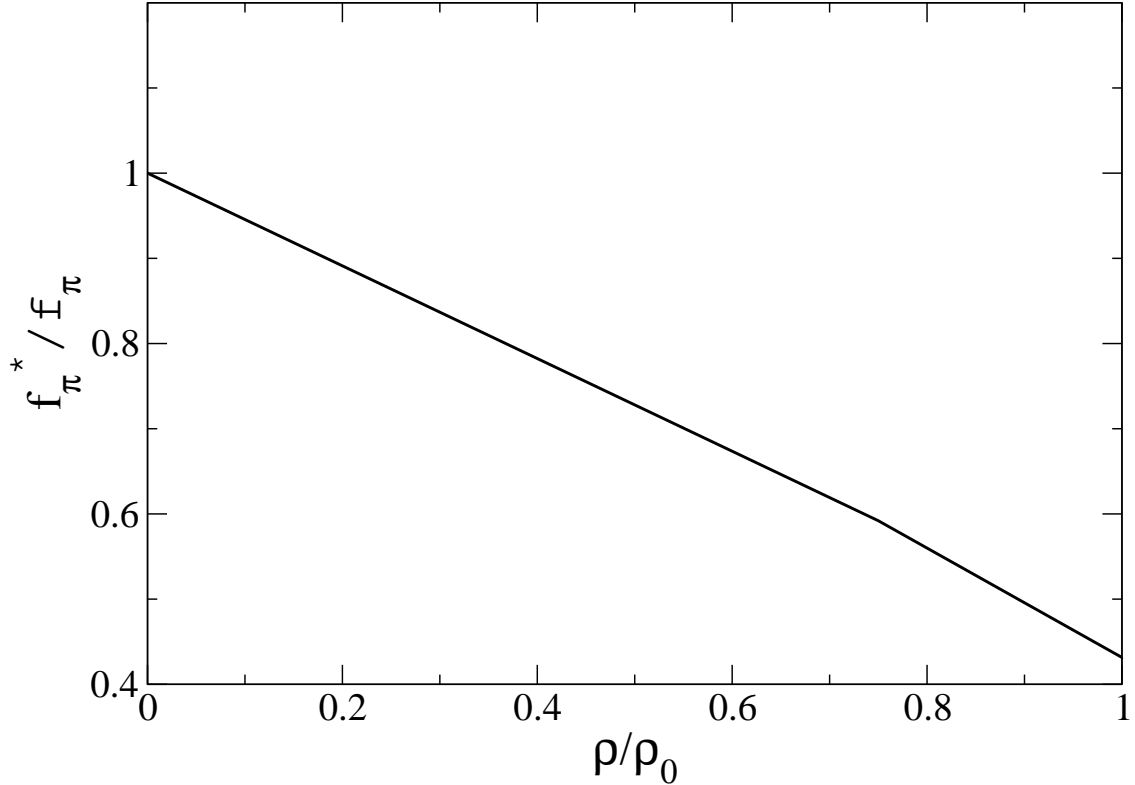}
\caption{Ratio of the in-medium to vacuum pion decay constant, $f^*_\pi / f_\pi$, associated
with the light-front time component, versus nuclear density.}
\label{Fig:fpi}
\end{center}
\end{figure*}

Next, we show in Fig.~\ref{Fig:fpi} the ratio of the in-medium to vacuum
pion decay constant, $f^*_\pi/f_\pi$, versus nuclear density,
associated with the light-front time component.
The result shows that $f^*_\pi$ decreases as nuclear density increases.
This is consistent with the empirical findings
based on the pionic-atom experiment~\cite{Kienle}, which yield $(f^*_\pi/f_\pi)^2 \simeq 0.64$
(associated with the time component) at density $\rho = 0.17$ fm$^{-3}$,
while our result yields a larger reduction.

Finally, we summarize in Table~\ref{Tab:summary} some properties of the pion in symmetric nuclear matter.
In particular, it is interesting to focus on the last column for $\eta^*$, which is the probability
of the valence component of the pion in symmetric nuclear matter, Eq.~(\ref{Eq:vcompo}).
As nuclear density increases, the probability of the valence
component in the pion is enhanced, which is again the effect of the decreasing in
the effective quark mass. This makes the quarks freer to move inside the pion than
the heavier valence quarks. This effect has the same origin as the increase of
the pion charge radius in nuclear matter discussed above.

\begin{table}[htb]
\begin{center}
\caption{
Summary of in-medium pion properties. $\eta^*$ is calculated via Eq.~(\ref{prob1}), the probability of the valence
component in the pion.
}
\label{Tab:summary}
\vspace*{3mm}
\begin{tabular}{|c|c|c|c|c|}
\hline
$\rho/\rho_0$  & $m^*_q$~[MeV] & $f^*_{\pi}$~[MeV] & $<r^{*2}_{\pi}>^{1/2}$~[fm] & $\eta^*$\\
\hline
~0.00     & ~220  & ~93.1   & ~0.73   & ~0.782 \\
~0.25  &  ~179.9  & ~80.6   & ~0.84   & ~0.812 \\
~0.50  &  ~143.2  & ~68.0   & ~1.00   & ~0.843 \\
~0.75  &  ~109.8  & ~55.1   & ~1.26   & ~0.878 \\
~1.00  &  ~79.5   & ~40.2   & ~1.96   & ~0.930 \\
\hline
\end{tabular}
\end{center}
\end{table}

In the present light-front model, it is not straightforward to discuss the in-medium
quark condensate and Gell-Mann-Oakes-Renner (GMOR) relation~\cite{GMOR},
as we use constituent quarks with $m_q = 220$~MeV in vacuum.
However, for illustration, we attempt to analyze
the GMOR-like relation and discuss the quark condensates
within the present approach. The difference with the usual GMOR relation is
that the pion decay constant in vacuum, $f_\pi$,
and in-medium, $f^*_\pi$, are calculated using constituent quark masses instead of
current quark masses. Keeping this in mind,
the GMOR-like relation in vacuum and in-medium may be written by,
\begin{eqnarray}
m^2_\pi f^2_\pi &=& - 2 m_q <\overline{q} q>,
\label{GMOR}\\
m^{*2}_\pi f^{*2}_\pi &=& - 2 m^*_q <\overline{q} q>^* \ .
\label{GMORm}
\end{eqnarray}
The ratio of the in-medium to vacuum quark condensates in the present approach may be estimated as,
\begin{equation}
\frac{<\overline{q} q>^*}{<\overline{q} q>} = \frac{m_q}{m^*_q} \frac{m^{*2}_\pi f^{*2}_\pi}{m^2_\pi f^2_\pi}
\simeq \frac{m_q}{m^*_q} \frac{f^{*2}_\pi}{f^2_\pi} \ .
\label{qconr}
\end{equation}
At normal nuclear matter density, $\rho_0$ (0.15 fm$^{-3}$), the ratio gives $\simeq 0.52$
using Table~\ref{Tab:summary}. This implies a larger reduction in ``quark condensate'' compared to
the value $0.67 \pm 0.06$ extracted in Ref.~\cite{Kienle} at a density 0.17 fm$^{-3}$ (their value for
the normal nuclear matter density). This feature may also be understood from the larger reduction in
$(f^*_\pi/f_\pi)^2$ in our approach compared with that obtained in Ref.~\cite{Kienle}.

\section{Summary and Discussions}
\label{summary}

We have studied the modifications of the pion properties in symmetric nuclear matter based on
the constituent quark model of the pion on the light front, where the pion model reproduces
well experimental data in vacuum. In order to incorporate the nuclear
many-body effects on an equal footing, i.e. with the quark degrees of freedom,
we have employed the QMC model.
We have made use of the in-medium quark properties obtained in the QMC model as input for
the constituent up and down quarks in the pion to study the in-medium modifications of
the pion properties. The in-medium quarks in the pion contain the information
of nuclear many-body dynamics, the nuclear Fermi momentum and nuclear saturation properties,
which are consistent at the level of the Hartree mean-field approximation.
Although this study is of exploratory nature, we believe that it constitutes an
advance in the treatment of the quarks confined in the pion in a nuclear medium.

With regard to the pion properties in symmetric nuclear matter, we have presented
the in-medium electromagnetic form factor, charge radius and decay constant up to normal
nuclear matter density, based on the plus (light-front time) component of the corresponding
light-front current.  Our results indicate a faster falloff of the elastic form factor with
increasing nuclear density, and consequently an increase of the pion charge radius.

Moreover, we have computed the in-medium pion decay constant, which is again associated
with the light-front time component.
We have shown that the decay constant decreases as nuclear density increases,
which is consistent with empirical findings based on the analysis of the
pionic-atom data.
The corresponding ratio, $f^*_\pi/f_\pi$, obtained in the present approach is smaller,
or equivalently, the reduction of $f_\pi$ is larger.
However, we should mention that in the empirical extraction an uncertainty in the in-medium
pion-mass shift exists, from which the value of the pion decay constant reduction is extracted.
(And we stress once again that in our case $f^*_\pi$ is the light-front time component.)

Concerning the valence quark probability in the pion, our result shows this probability increases
with increasing nuclear density. We interpret this in terms of the decrease
in the effective constituent quark masses in the pion,
which allows for a larger kinematical distribution of the quarks
within the pion, and in turn results in the increase of the valence probability.
The same reasoning holds for the increase of the pion charge radius.

We have also estimated the in-medium quark condensate using the Gell-Mann-Oakes-Rener-like relation,
and obtained the reduction of the in-medium quark condensate relative to that in vacuum.
However, the reduction is larger than that from the pionic-atom data analysis.
Most likely, this is due to the large constituent quark
masses used in the pion model.

In future, the present approach may be extended to the kaon, $D$-, $\rho$- and $\omega$-mesons.
Alternatively, we can treat the in-medium effects on the quark's mass function
by means of a Dyson-Schwinger equation with finite density and incorporate them in
the Bethe-Salpeter equation for the bound states.

\section*{Acknowledgments}
This work was partially supported by the Brazilian agencies CNPq and
FAPESP. The work of K. Tsushima was also supported by the Brazilian Ministry of Science,
Technology and Innovation (MCTI-Brazil), and
Conselho Nacional de Desenvolvimento Cient{\'i}fico e Tecnol\'ogico
(CNPq), project 550026/2011-8.


\end{document}